\documentclass[a4paper,10pt,twoside]{cpc-hepnp}
\usepackage{upgreek,fancyhdr}
\usepackage{multicol}
\usepackage{graphicx}
\usepackage{booktabs}
\usepackage{amssymb,bm,mathrsfs,bbm,amscd}
\usepackage[tbtags]{amsmath}
\usepackage{lastpage}
\usepackage[english]{babel}

\begin{document}

\fancyhead[c]{\small Chinese Physics C~~~Vol. xx, No. x (2020) xxxxxx}
\fancyfoot[C]{\small 010201-\thepage}
\footnotetext[0]{Received 31 June 2020}

\title{A new method to test the cosmic distance duality relation using the strongly lensed gravitational waves\thanks{Supported by the National Natural Science Fund of China (Grant Nos. 11603005 and 11775038).}}

\author{%
Hai-Nan Lin$^{1;1)}$\email{linhn@cqu.edu.cn}
\quad Xin Li$^{1;2)}$\email{lixin1981@cqu.edu.cn}
}
\maketitle

\address{%
$^1$ Department of Physics, Chongqing University, Chongqing 401331, China\\
}

\begin{abstract}
  We propose a new method to test the cosmic distance duality relation using the strongly lensed gravitational waves. The simultaneous observations of image positions, the relative time delay between different images, the redshift measurements of lens and source, together with the mass modelling of the lens galaxy, provide the angular diameter distance to the gravitational wave source. On the other hand, from the observation of gravitational wave signals the luminosity distance to the source can be obtained. This is to our knowledge the first method to simultaneously measure both the angular diameter distance and luminosity distance from the same source. Thus, the strongly lensed gravitational waves provide a unique way to test the cosmic distance duality relation. With the construction of the third generation gravitational detectors such as the Einstein Telescope in the future, it is possible to test the cosmic distance duality relation at an accuracy of several percent.
\end{abstract}



\footnotetext[0]{\hspace*{-3mm}\raisebox{0.3ex}{$\scriptstyle\copyright$}2019
Chinese Physical Society and the Institute of High Energy Physics
of the Chinese Academy of Sciences and the Institute
of Modern Physics of the Chinese Academy of Sciences and IOP Publishing Ltd}%

\begin{multicols}{2}

\section{Introduction}\label{sec:introduction}

Due to the expansion of the universe there are several ways to define distance in cosmology, among which the luminosity distance and angular diameter distance are two important definitions. The former is defined by the fact that for an object with fixed luminosity, the measured flux is inversely proportional to the square of distance to the source, while the latter is defined as the ratio of an object's physical transverse size to its angular size. The cosmic distance duality relation (DDR) correlates the luminosity distance to the angular diameter distance by $D_L(z)=(1+z)^2D_A(z)$ \cite{Etherington:1933,Etherington:2007}. The standard DDR holds true in any metric theory of gravity such as the general relativity, as long as the photons travel along null geodesics and the photon number is conserved during the propagation \cite{Ellis:1971,Ellis:2007}. The violation of DDR may be caused by e.g. the extinction of photon by intergalactic dust \cite{Corasaniti:2016}, the coupling of photon with other particles \cite{Bassett:2003vu}, the variation of fundamental constants \cite{Ellis:2013}. The DDR is a fundamental relation in the standard cosmological model. Any violation of DDR would imply that there are new physics beyond the standard cosmological model. Therefore, testing the validity of DDR is of great importance. In fact several works have been devoted to test the DDR \cite{Holanda:2010vb,Piorkowska:2011nhd,Yang:2013coa,Costa:2015lja,Holanda:2016msr,Ma:2016bjt,Holanda:2016zpz,Li:2018,Hu:2018yah,Lin:2018mdj,Liao:2019xug}.

The method to test DDR is simple: just measure the angular diameter distance ($D_A$) and luminosity distance ($D_L$) to the same redshift, then compare these two distances to see if the DDR is valid or not. However, in practice this is not a trivial thing. Although $D_A$ and $D_L$ can be measured in several ways independently, it is difficult to simultaneously measure $D_A$ and $D_L$ from the same object. The usual way is to measure $D_A$ and $D_L$ from different objects locating at different positions. For example, the luminosity distance can be measured from the type-Ia supernovae standard candles \cite{Riess:1998mnb,Perlmutter:1998np}, the gravitational waves standard sirens \cite{Abbott:2016ajs,Abbott:2017lpn}. The angular diameter distance can be measured e.g. from the BAO signals in the galaxy spectrum \cite{Beutler:2011hx,Anderson:2013zyy}, the Sunyaev-Zel'dovich effect of galaxy clusters \cite{Filippis:2005,Bonamente:2006ct}, the angular size of ultra-compact radio sources \cite{Jackson:2006bg}, the strong gravitational lensing system \cite{Liao:2016uzb}.

One main problem of the above method is that $D_A$ and $D_L$ are measured from different objects locating at different redshifts and sky positions. To test DDR one must apply special techniques such as interpolations and Gaussian processes so that $D_L$ and $D_A$ can be compared at the same redshift. Although the standard DDR only involves the distances but not the directions on the sky, if the universe is anisotropic (caused by e.g. the inhomogeneous and anisotropic distribution of matter) it is unreasonable to test DDR using $D_A$ and $D_L$  measured from objects locating at different directions \cite{Li:2018tlj}. The ideal way to avoid this problem is of course to measure $D_A$ and $D_L$ from the same object, and then directly compared these two distances. So is there any way to measure $D_A$ and $D_L$ from the same source? We will show in the following that the strongly lensed gravitational waves can satisfy our requirement.

In this paper, we propose a new method to test the DDR using the strongly lensed gravitational waves. The spectroscopic observation of redshifts of lens and source, the photometric observation of lens galaxy, together with the observations of GW image positions and the relative time delay between images, give both the angular diameter distance and luminosity distance to the GW source. Thus the strongly lensed GWs provide a unique way to test the DDR using a signal source.
The structure of this paper is arranged as follows: The method is described in section 2, together with some related discussions in section 3. Finally, a short summary is given in section 4.

\section{Methodology}\label{sec:method}

We consider the situation where a GW event originates from the coalescence of compact binary system (e.g. NS-NS binary and NS-BH binary) is strongly gravitationally lensed by a foreground galaxy. We also assume that the lens galaxy is modeled as a singular isothermal sphere. With this configuration, two images appear at the angular positions $\theta_1$ and $\theta_2$ with respect to the lens position. The Einstein radius $\theta_E=|\theta_1-\theta_2|/2$ is given by \cite{Mollerach:2002}
\begin{equation}\label{eq:thetaE}
  \theta_E=\frac{4\pi\sigma_{\rm SIS}^2D_A(z_l,z_s)}{c^2D_A(z_s)},
\end{equation}
where $\sigma_{\rm SIS}$ is the velocity dispersion of the lens galaxy, $D_A(z_s)$ and $D_A(z_l,z_s)$ are the angular diameter distances from the observer to source and from the lens to source, respectively. If the angular resolution of the GW detector is high enough such that the angular positions of the two images can be well measured so the Einstein radius is preciously known, and if the velocity  dispersion of the lens galaxy is measured independently, then we can obtain the distance ratio
\begin{equation}\label{eq:Dls2Ds}
  R_A\equiv\frac{D_A(z_l,z_s)}{D_A(z_s)}=\frac{c^2\theta_E}{4\pi\sigma_{\rm SIS}^2}.
\end{equation}

On the other hand, two images of GW propagating along different paths will cause relative time delay, which is given by \cite{Mollerach:2002}
\begin{equation}\label{eq:timedelay}
  \Delta t=(1+z_l)\frac{D_{\Delta t}}{c}\Delta\phi,
\end{equation}
where
\begin{equation}\label{eq:D_timedelay}
  D_{\Delta t}\equiv\frac{D_A(z_l)D_A(z_s)}{D_A(z_l,z_s)}=\frac{c}{1+z_l}\frac{\Delta t}{\Delta\phi}
\end{equation}
is the so-called time-delay distance, and
\begin{equation}\label{eq:Fermi_potential}
  \Delta\phi=\frac{(\theta_1-\beta)^2}{2}-\Psi(\theta_1)-\frac{(\theta_2-\beta)^2}{2}+\Psi(\theta_2)
\end{equation}
is the difference of Fermat potential of the lens galaxy calculated at the image positions, $\Psi(\theta)$ is the rescaled projected gravitational potential of the lens galaxy. For the singular isothermal spherical lens, $\Psi(\theta)=\theta_E|\theta|$. If the gravitational potential of the lens galaxy can be well measured from the photometric and dynamical observations such that the Fermat potential can be calculated, and if the spectroscopic redshift of the lens galaxy is precisely known, then the time-delay distance can be determined from the observed time delay between two GW images.

In a spatially flat universe, the comoving distance is related to the angular diameter distance by $r(z_s)=(1+z_s)D_A(z_s)$, $r(z_l)=(1+z_l)D_A(z_l)$, $r(z_l,z_s)=(1+z_s)D_A(z_l,z_s)$, where the comoving distance from lens to source is simply given by $r(z_l,z_s)=r(z_s)-r(z_l)$. Therefore, the angular diameter distance from lens to source reads
\begin{equation}\label{eq:Dls}
  D_A(z_l,z_s)=D_A(z_s)-\frac{1+z_l}{1+z_s}D_A(z_l).
\end{equation}
Equations (\ref{eq:Dls2Ds})(\ref{eq:D_timedelay})(\ref{eq:Dls}) uniquely solve for $D_A(z_l)$, $D_A(z_s)$ and $D_A(z_l,z_s)$.

What we are interested in is the distance from observer to source, which reads
\begin{equation}\label{eq:DA_zs}
  D_A(z_s)=\frac{1+z_l}{1+z_s}\frac{R_AD_{\Delta t}}{1-R_A},
\end{equation}
where $R_A$ and $D_{\Delta t}$ are given by equations (\ref{eq:Dls2Ds}) and (\ref{eq:D_timedelay}), respectively. Using the error propagating formulae we obtain the uncertainty on $D_A(z_s)$,
\begin{equation}\label{eq:error_DA_zs}
  \frac{\delta D_A(z_s)}{D_A(z_s)}=\sqrt{\left(\frac{\delta R_A}{R_A(1-R_A)}\right)^2+\left(\frac{\delta D_{\Delta t}}{D_{\Delta t}}\right)^2},
\end{equation}
where \begin{equation}\label{eq:error_RA}
  \frac{\delta R_A}{R_A}=\sqrt{\left(\frac{\delta\theta_E}{\theta_E}\right)^2+4\left(\frac{\delta\sigma_{\rm SIS}}{\sigma_{\rm SIS}}\right)^2},
\end{equation}
and
\begin{equation}\label{eq:error_Dt}
  \frac{\delta D_{\Delta t}}{D_{\Delta t}}=\sqrt{\left(\frac{\delta \Delta t}{\Delta t}\right)^2+\left(\frac{\delta\Delta\phi}{\Delta\phi}\right)^2}.
\end{equation}
If the observables ($z_l$, $z_s$, $\Delta t$, $\Delta \phi$, $\theta_E$, $\sigma_{\rm SIS}$) are measured, then $D_A(z_s)$ and its uncertainty can be obtained using equations (\ref{eq:DA_zs})--(\ref{eq:error_Dt}).

The luminosity distance to the source $D_L(z_s)$ can be inferred directly from the GW signals \cite{Abbott:2016ajs,Abbott:2017lpn}. As standard sirens, GWs can provide luminosity distance model-independently, thus are widely used as cosmological probes \cite{Chen:2017rfc,Cai:2017aea,Lin:2018azu,Chang:2019dkj}. $D_L(z_s)$ is inversely proportional to the amplitude of spacetime strain in the Fourier space, $D_L(z_s)\propto 1/\mathcal{A}$. Due to the degeneracy between $D_L(z_s)$ and the inclination angle of the binary's orbital plane, the uncertainty on $D_L(z_s)$ maybe very large. However, if the GW event is accompanied by a short gamma-ray burst (GRB), then due to the beaming of GRB outflow we can assume that the inclination angle is small, hence the degeneracy breaks. In this case the uncertainty on $D_L(z_s)$ can be estimated as \cite{Sathyaprakash:2009xt,Cai:2017sby}
\begin{equation}\label{eq:error_DL}
  \frac{\delta D_L(z_s)}{D_L(z_s)}=\sqrt{\left(\frac{2}{\rho}\right)^2+(0.05z_s)^2},
\end{equation}
where $\rho$ is the signal-to-noise ratio (SNR) of the detector's response to GW signal, and the $0.05z_s$ term represents the uncertainty arising from weak lensing effect caused by the matter distribution along the line-of-sight.

Note that $D_L(z_s)$ directly inferred from the GW signals is not the true luminosity distance. This is because $D_L(z_s)$ is inversely proportional to the amplitude of GW strain, while the latter is magnified by the lensing effect. For the singular isothermal spherical lens the magnification is given by $\mu_\pm =1\pm\theta_E/\beta$, where $\beta$ is the actual position of the source, and ``$\pm$" represent the first and second images, respectively. The actual position of the source $\beta$ can be determined through deep photometric imaging, $\beta/\theta_E=(F_+-F_-)/(F_++F_-)$, where $F_\pm$ are the photometric flux of two images. Given the magnification factor determined from the photometric observations, the true distance can be obtained by $D_L^{\rm true}=\sqrt{\mu_\pm}D_L^{\rm obs}$. The uncertainty of $\mu_\pm$ will propagate to $D_L$. Therefore, the final uncertainty on $D_L(z_s)$ is given by
\begin{equation}\label{eq:error_on_dL}
  \frac{\delta D_L(z_s)}{D_L(z_s)}=\sqrt{\left(\frac{2}{\rho}\right)^2+(0.05z_s)^2+\frac{1}{4}\left(\frac{\delta\mu_\pm}{\mu_\pm}\right)^2}.
\end{equation}
Since the magnification is derived directly from the photometric fluxes of two images, the uncertainty on $D_L(z_s)$ is uncorrelated with that on $D_A(z_s)$. If two GW images are observed, the distance inferred from different images can be used to crosscheck with each other and the uncertainty on $D_L(z_s)$ can be further reduced.

Given that both the angular diameter distance and luminosity distance are measured, the DDR can be directly tested. We define the possible deviation from the standard DDR as
\begin{equation}\label{eq:Delta}
  \Delta=\frac{D_L-(1+z)^2D_A}{D_L}.
\end{equation}
If the uncertainty on the luminosity distance is uncorrelated with that on the angular diameter distance, then the uncertainty on $\Delta$ is given by
\begin{eqnarray}\label{eq:uncertainty}
  \delta\Delta&=&(1+z)^2\sqrt{\frac{(\delta{D_A})^2}{D_L^2}+\frac{D_A^2(\delta{D_L})^2}{D_L^4}}\\ \nonumber
  &=&\sqrt{\left(\frac{\delta{D_A}}{D_A}\right)^2+\left(\frac{\delta{D_L}}{D_L}\right)^2},
\end{eqnarray}
where in the last equality we have assumed that the violation of DDR, if really exists, is very small. If DDR is valid, $\Delta$ should be consistent with zero. Any deviation of equation (\ref{eq:Delta}) from zero would imply the violation of DDR.

In summary, both $D_A(z_s)$ with $D_L(z_s)$ can be measured from the strongly lensed GW system. This provides a unique way to simultaneously measure the angular diameter distance and luminosity distance from the same object. This method is independent of cosmological models, except the assumption that the universe is spatially flat. Therefore, the strongly lensed GW provides a model-independent tool to test the DDR.

\section{Discussions}\label{sec:discussion}

Although the idea proposed here seems to be theoretically promising, in practice there are many challenges. The biggest challenge is the identification of the lensed GW signals. For a typical lensing system the angular separation between two images is at the order of arc seconds. This is far beyond the angular resolution of the in-running GW detectors such as LIGO and Virgo. Even for the in-planed third generation GW detectors such as Einstein Telescope and Cosmic Explorer, the situation is not more optimistic. However, the localization capability of a network of third generation GW detectors is expected to be precise enough to identify the host galaxy \cite{Zhao:2017cbb}. If the host galaxy can be identified and electromagnetic counterparts can be observed, then different images can be distinguished through photometric observations. If we assume that light and GW propagate along the same null geodesics, then the positions of photometry images overlap with the positions of GW images. The GW is a transient event which lasts at most several seconds, while the time delay between different images is typically at the order of magnitude of several months or even several years. This makes the observation of GW lensing much more difficult than the observation of regular lensing, such as the lensing of quasar and supernova. The GW detectors must keep running to ensure that both images can be recorded.

Up to now, at least more than one hundred strong gravitational lensing systems in which quasar acts as the source have been found, see e.g. the catalog compiled by \cite{Cao:2015qja}. The redshift of the source is usually in the range from $z_s\sim0.5$ to $z_s\sim3.5$. If we assume that the lensed GW sources fall into the similar redshift range to the quasars, it is far beyond the effective detection range of LIGO and Virgo. However, this distance is reachable by the in-planed Einstein Telescope, which is designed to be able to detect GW events up to $z\sim 5$. It is expected that about $10^3\sim 10^7$ GW events can be detected by the Einstein Telescope per year \cite{ET}, among which several events may be strongly lensed by a foreground galaxy. It is optimistically estimated that about $50-100$ strongly lensed GW events per year can be observed by the Einstein Telescope \cite{Biesiada:2014kwa}. With the space-based detectors such as the Big Bang Observer \cite{Cutler:2006aks}, the detection rate is expected to be much higher. Among these lensed GW events, several of them are expected to be produced by the coalescence of NS-NS binary or NS-BH binary, which may be accompanied by electromagnetic counterparts, hence the redshift can be measured independently. In addition, the redshift range also fall into the effective detection range of some existing gamma-ray burst detectors such as the Fermi satellite, so the joint observations of GWs and electromagnetic counterparts are possible. Thus, despite the big challenges, there is still great chance to detect the strongly lensed GW events in the near future.

The uncertainty of $D_A(z_s)$ mainly comes from the measurements of $\Delta\phi$, $\theta_E$ and $\sigma_{\rm SIS}$. It is showed that \cite{Liao:2017ioi} in the strongly lensed GW systems the accuracy of $\Delta\phi$ can be improved by a factor of five compared to the strongly lensed quasar systems because the GW signals do not suffer from the bright AGN contamination from the lens galaxy. So we follow \cite{Liao:2017ioi} and assume $0.6\%$ uncertainty on $\Delta\phi$. The accuracy of $\theta_E$ is expected to be at $\sim 1\%$ level in the future LSST survey \cite{Cao:2019kgn}. According to the quasar lensing systems compiled by \cite{Cao:2015qja}, the measured uncertainty of velocity dispersion of the lens galaxy is at the order of $\sim 10\%$, with the best accuracy $\sim 3\%$. Due to the improvements of observation technique, it is expected that the uncertainty can be further reduced in the future. Here we follow \cite{Cao:2019kgn} and assume the uncertainty of $\sigma_{\rm SIS}$ at the level of $5\%$. If the host galaxy is identified, the redshifts of lens and source can be measured precisely in the spectroscopic way so their uncertainties are negligible. Due to the transient property of GW event, the time delay between two GW images can also be precisely measured with a negligible uncertainty. It is shown that the redshift of the strongly lensed GW source has a sharp peak near $z_s\sim 2$ \cite{Ding:2015uha}, and the median value of redshift of the lens is $z_l\sim 0.8$ \cite{Cao:2019kgn}. With the expected accuracy, for a typical lens system at $z_l\sim 0.8$ and $z_s\sim 2$, the uncertainty on $D_A(z_s)$ is estimated according to equation (\ref{eq:error_DA_zs}) to be $\delta D_A/D_A\sim 20\%$. In fact, the uncertainty mainly arises from the error on $\sigma_{\rm SIS}$. To reduce the uncertainty more precise measurement of $\sigma_{\rm SIS}$ is needed.

The uncertainty of $D_L(z_s)$ mainly comes from three aspects: the SNR of GW signal recorded by the detector, the week lensing effect caused by the matter along the light path, the uncertainty of the magnification factor. To ensure the significance, here we require that the SNR is at least $\rho\gtrsim 16$, compared to the usual criterion $\rho\gtrsim 8$. For a GW source locating at $z_s\sim 2$, the uncertainty cause by the weak lesing effect is about $10\%$. Hence the uncertainty of $D_L(z_s)$, according to equation (\ref{eq:error_DL}), is at the level of $16\%$. For a source locating at redshift higher than 2, the uncertainties on the luminosity distance is dominated by the weak lensing term, so the accuracy may be worse than our rough estimate. However, if two GW images are observed, both images can be used to determine the luminosity distance, hence the accuracy can be improved by a factor of $\sqrt{2}$. Therefore, it is still possible to measure $D_L(z_s)$ with an accuracy better than $16\%$ for high-redshift events. The determination of magnification factor $\mu_\pm$ may be highly uncertain due to the contamination of image flux by the foreground lensing galaxy. Here we follow \cite{Cao:2019kgn} and assume $\sim 20\%$ uncertainty on $\mu_\pm$. Taking the uncertainty of magnification factor into consideration, the final uncertainty on $D_L(z_s)$ is at $\sim 20\%$ level.

Based on the above discussions, we may expect that in the near future, although is still challenging, it is possible to measure both $D_A$ and $D_L$ to the GW source with an accuracy better than $\sim 20\%$. With this accuracy, a single strongly lensed GW event could, according to equation (\ref{eq:uncertainty}), constrain DDR at the level of $~\sim 28\%$. If $\sim 100$ events with similar accuracy are detected, the DDR can be constrained at $~\sim 3\%$ level. Although the accuracy of this method within the present technique may be not as competitive as the methods mentioned in the introduction, this is the first way to test DDR using a signal source.

Here we just use the singular isothermal spherical (SIS) model as an example to show how our method works. In actual case the lens model is generally more complex than the simple SIS model. If the matter distribution of the lens deviates from the simple SIS model, for example the more general power-law model, most of our formulae still work, except the change of Einstein radius (equation \ref{eq:thetaE}) and the concrete form of Fermi potential (equation \ref{eq:Fermi_potential}). Besides, some other uncertainties arise from the lens parameters, e.g. the index of the power-law model. What one needs to do is to accurately model the mass distribution of the lens galaxy, in regardless of its concrete form. It is shown that the redshift of the lens galaxy in strongly lensed GW systems is usually bellow $z\sim 1$ \cite{Cao:2019kgn,Liao:2019xug}. Actually this is in the similar redshift range of the lens galaxies in the quasar lensing systems. At this redshift range, the dispersion velocity and mass distribution of the lens galaxy is expected to be measured with an acceptable accuracy in the future LSST survey using the photometric and spectroscopic data.

The method proposed here is free of cosmological model, but we assume that the cosmos is spatially flat. This assumption is reasonable because the observations on the cosmic microwave background radiation by the Planck satellite show that the data is well consistent with a flat universe \cite{Ade:2016xua,Aghanim:2018eyx}. If the universe is non-flat, the exact value of the space curvature should be known in order to calculate the distance from the lens to source. Then equations (\ref{eq:Dls}) and (\ref{eq:DA_zs}) should be modified which depend on the curvature. Very recently, some works find moderate evidence for a closed universe \cite{DiValentino:2019qzk,Handley:2019tkm}. Even if the universe is really non-flat, due to the smallness of the space curvature, in the local universe where the strongly lensed GW events can be detected ($z\lesssim 5$) the distance measures is not strongly affected. Besides, the uncertainty of the lens modelling and the GW signal is much larger than the uncertainty caused by the possible non-flatness of the universe. Therefore, our analysis does not strongly depend on the specific value of the space curvature.

It should be noted that all of our calculations are based on the framework of general relativity. The violation of general relativity may, but not necessarily cause the violation of DDR. If the future data show strong evidence for the violation of DDR, there are still some interpretations. The violation of DDR may be caused by some reasons, e.g., the dust extinction, the coupling of photon or graviton with some unknown particles, the photon or graviton being massive, etc. The method proposed here can only test if DDR is valid or not. To interpret the reason for the violation of DDR, however, requires further investigations.

\section{Summary}\label{sec:summary}

In this paper, we proposed a new method to test the DDR using the strongly lensed GW event. The photometric and spectroscopic observations of the source and lens galaxies, combined with the GW observation provides a unique way to measure both the angular diameter distance and luminosity distance to the GW source. This is to our knowledge the first method which can measure both the angular diameter distance and luminosity distance from the same object up to high redshifts. Although this method is beyond the present-day observational technology, we couldn't exclude the possibility that it can be put into practice with the construction of the third generation GW detectors in the near future.



\end{multicols}

\vspace{-1mm}
\centerline{\rule{80mm}{0.1pt}}
\vspace{2mm}

\begin{multicols}{2}

\end{multicols}

\end{document}